\documentclass[aps,prd,superscriptaddress,nofootinbib,amsmath,amsfonts,preprintnumbers,groupedaddress,showpacs,9pt,english]{revtex4-2}
\usepackage{amsmath}
\usepackage{amssymb}
\usepackage{babel}
\usepackage{wrapfig}
\usepackage{cancel}

\usepackage{graphicx}
\graphicspath{{figures/}}
\usepackage{amsmath,amssymb} 
\usepackage{mathrsfs}    
\usepackage[colorlinks,citecolor=blue,urlcolor=blue,linkcolor=blue]{hyperref}
\usepackage[figtopcap]{subfigure}
\usepackage{color}
\usepackage{relsize,exscale}
\makeatletter
\newcommand{\beq}{\begin{equation}}
\newcommand{\eeq}{\end{equation}}
\newcommand{\bea}{\begin{eqnarray}}
\newcommand{\eea}{\end{eqnarray}}

\newcommand{\comm}[1]{}

\usepackage{array,multirow,graphicx}
\usepackage{dcolumn}
\usepackage{newlfont}
\usepackage{bm}

\usepackage{scalerel}
\usepackage{tikz}
\usetikzlibrary{svg.path}
\definecolor{orcidlogocol}{HTML}{A6CE39}
\tikzset{
  orcidlogo/.pic={
    \fill[orcidlogocol] svg{M256,128c0,70.7-57.3,128-128,128C57.3,256,0,198.7,0,128C0,57.3,57.3,0,128,0C198.7,0,256,57.3,256,128z};
    \fill[white] svg{M86.3,186.2H70.9V79.1h15.4v48.4V186.2z}
                 svg{M108.9,79.1h41.6c39.6,0,57,28.3,57,53.6c0,27.5-21.5,53.6-56.8,53.6h-41.8V79.1z M124.3,172.4h24.5c34.9,0,42.9-26.5,42.9-39.7c0-21.5-13.7-39.7-43.7-39.7h-23.7V172.4z}
                 svg{M88.7,56.8c0,5.5-4.5,10.1-10.1,10.1c-5.6,0-10.1-4.6-10.1-10.1c0-5.6,4.5-10.1,10.1-10.1C84.2,46.7,88.7,51.3,88.7,56.8z};}}
\newcommand\orcid[1]{\href{https://orcid.org/#1}{\mbox{\scalerel*{
\begin{tikzpicture}[yscale=-1,transform shape]
\pic{orcidlogo};
\end{tikzpicture}
}{|}}}}

\begin{document}

\title{A No-Go Theorem for Curvature Neutrality in $f(Q)$ Cosmology?}
\author{G.G.L. Nashed$^{1,2}$}\email{nashed@bue.edu.eg}
\author{Amare Abebe$^{2,3}$}\email{Amare.Abebe@nithecs.ac.za}

\affiliation{\\$^{1}$Centre for Theoretical Physics, The British University, P.O. Box 43, El Sherouk City, Cairo 11837, Egypt\\$^{2}$Center for Space Research, North-West University, Potchefstroom 2520, South Africa\\
$^{3}$National Institute for Theoretical and Computational Sciences (NITheCS), Potchefstroom 2520, South Africa}

\date{\today}
\begin{abstract}
In the framework of $f(Q)$ gravity, we investigate whether spatial curvature
can be made dynamically invisible in the cosmological background equations.
We consider open spatial sections, $k<0$, on an explicitly specified
homogeneous and isotropic symmetric-teleparallel connection branch. For
$k\neq0$, the coincident gauge cannot be imposed simultaneously with the
standard curved-FLRW coordinate form. Re-deriving the background equations
from the minisuperspace action, we show that curvature enters through
$x=H+\delta\sqrt{-k}/a$ and appears with inequivalent weights in the energy
and pressure equations. For $f\in C^3((0,\infty))$, we prove a branch-specific
no-go theorem: strict curvature independence for every scale factor and every
$k<0$ requires $f=\mathrm{const}$, which contains no metric kinetic term.
STEGR is not curvature-neutral, since its Friedmann equation retains the usual
$3k/a^2$ contribution. A weaker cancellation, imposed only on backgrounds
satisfying $\delta\sqrt{-k}/a=cH$, necessarily produces a coasting expansion
and yields $f(Q)=A Q^{(c+2)/2}+B$, or $f(Q)=A\ln Q+B$ when $c=-2$. For the
illustrative choice $c=-3$, the solution becomes
$f=\alpha_1/\sqrt Q+\beta_1$. Under the conventional positive-coupling
assumption $f_Q>0$, the required source violates the null energy condition.
Moreover, adding positive radiation or pressureless matter forces a
compensating negative-energy component at sufficiently early times. The weak
branch is therefore neither curvature-neutral in an invariant sense nor a
viable cosmological model. The principal result is the background-level
obstruction itself. Perturbative stability and gravitational-wave propagation
require a separate analysis including perturbations of the nontrivial affine
connection and are not established in this work.
\end{abstract}
\maketitle
\section{Introduction}

The discovery of the late-time accelerated expansion of the Universe has made
the origin of cosmic acceleration one of the principal questions of modern
cosmology
\cite{SupernovaSearchTeam:1998fmf,SupernovaCosmologyProject:1998vns,SDSS:2005xqv,Weinberg:2013agg}.
Besides introducing a dark-energy component, one may seek an explanation
through a modification of the gravitational interaction. A broad class of
theories has consequently been investigated, including $f(R)$ gravity,
scalar-tensor models, teleparallel and symmetric teleparallel formulations,
and vector-tensor theories
\cite{Clifton:2011jh,Nojiri:2010wj,Capozziello:2011et,DeFelice:2010aj,Sotiriou:2008rp,Brans:1961sx,Horndeski:1974wa,Cai:2015emx,Hehl:1994ue,BeltranJimenez:2017tkd,Heisenberg:2018vsk}.
Their common purpose is to describe cosmic acceleration through spacetime
geometry while remaining consistent with local and cosmological tests
\cite{Cai:2015emx,BeltranJimenez:2017tkd,Will:2014kxa,Khoury:2003rn,Joyce:2014kja}.

Teleparallel and symmetric teleparallel gravity provide two geometrically
distinct realizations of this program. In teleparallel gravity, curvature
vanishes and torsion describes the gravitational field. In symmetric
teleparallel gravity, both curvature and torsion vanish, while gravity is
encoded entirely in non-metricity
\cite{Hehl:1994ue,BeltranJimenez:2017tkd,Nester:1998mp,Hayashi:1979qx,Iosifidis:2018jwu,McCrea:1992wa,BeltranJimenez:2019esp,Bahamonde:2021gfp}. Together with the curvature
formulation of GR, these descriptions constitute the geometrical trinity of
gravity and show that equivalent dynamics may be represented by different
geometric objects.

The symmetric teleparallel equivalent of general relativity (STEGR) is the
linear theory in this framework. It reproduces the Einstein equations up to a
boundary term and allows the affine connection to vanish in an appropriate
coordinate system, known as the coincident gauge
\cite{BeltranJimenez:2017tkd,BeltranJimenez:2019esp,Tomonari:2024vij,BeltranJimenez:2019tme,BeltranJimenez:2019odq,Bahamonde:2022zgj,Blixt:2023kyr,Capozziello:2021pcg,Capozziello:2023vne,Paliathanasis:2023kqs}.
This property can simplify the field equations considerably. Its direct
generalization is $f(Q)$ gravity, in which the gravitational Lagrangian is an
arbitrary function of the non-metricity scalar $Q$
\cite{BeltranJimenez:2019tme,Bahamonde:2022zgj}.

Cosmological applications of $f(Q)$ gravity include background evolution,
late-time acceleration, and observational constraints
\cite{Lazkoz:2019sjl,Shi:2023kvu,Narawade:2023rip,Koussour:2023rly,Sultanaa:2025ooz,Garg:2025vyo};
matter-coupled extensions of the theory have also been studied
\cite{Hazarika:2024alm}.
Nonzero-curvature backgrounds and their connection dependence have also been
studied using self-similar solutions, dynamical systems, and coincident-gauge
coordinates \cite{Dimakis:2022wkj,Paliathanasis:2023raj,Jensko:2024bee}.
The degree-of-freedom content and stability of a cosmological solution are,
however, connection-branch dependent and require a perturbative analysis
beyond the background equations
\cite{Heisenberg:2023wgk}.

In a parallel development, a flat-like universe (FLU) prescription was proposed
in $f(T)$ gravity to construct backgrounds whose evolution mimics that of a
spatially flat universe despite nonvanishing spatial curvature
\cite{ElHanafy:2014tzj,ElHanafy:2015jbo,ElHanafy:2014efn,Nashed:2014lva}.
That construction isolates an explicit curvature contribution, sets a
coefficient to zero, and supplements this condition with a further ansatz used
to determine the model and background. It is therefore a particular
background/model-selection prescription, not the strong fixed-history
criterion introduced below.

The corresponding construction in $f(Q)$ gravity, which we denote by CNQ, is
more subtle. On the nontrivial FLRW-compatible connection branch, spatial
curvature enters through several independent geometric structures. It appears
in mixed Hubble-curvature terms and implicitly through $Q$ and its derivatives.
This structure prevents a direct identification of the $FU_T$ coefficient
condition with the criterion used here and places stronger restrictions on the
form of the Lagrangian.

An especially relevant result is the coincident-gauge construction of the
unique spatially curved FLRW branch in Ref.~\cite{Jensko:2024bee}. For
$k<0$, imposing the connection consistency condition off shell and
independently of the choice of $f$ gives, after translating sign conventions,
$Q=-6(H+\delta\sqrt{-k}/a)^2$ as used below and yields a correspondence with the
background equations of metric teleparallel $f(T)$ gravity. Our no-go theorem
does not dispute that correspondence. It asks a different question: whether,
at fixed $a$, $H$, and their time derivatives, both source functions are
independent of the curvature scale. Nor do we assume that the earlier FLU
construction satisfies this stronger definition.

The purpose of the present study is to test this strong source-map notion of
curvature neutrality in $f(Q)$ gravity on the FLRW-compatible connection for
open spatial sections. We formulate the requirement as complete curvature
independence at fixed kinematic history and prove that it admits only
$f=\mathrm{const}$. Linearizing
the Lagrangian does not evade the obstruction: STEGR, and hence GR, retains
the standard curvature term and also fails the condition. Strict CNQ is
therefore not a selection principle for a nondegenerate $f(Q)$ model on this
branch. The novelty claimed here is this no-go result for the stronger
criterion, rather than the construction of the curved branch itself.

We then examine a weaker cancellation imposed on one constrained background.
It produces a one-parameter family of Lagrangians but necessarily forces a
coasting expansion. We derive the solution and determine the matter source
required to support it. For the conventional sign $f_Q>0$, the total source
violates the null energy condition. If positive radiation or pressureless
matter is included, the residual component becomes negative at sufficiently
early times. We therefore present this weak branch as a diagnostic of the
obstruction and not as a viable cosmological model.

The paper is organized as follows. In Sec.~\ref{sec:1}, we review symmetric
teleparallel gravity and introduce the $f(Q)$ field equations, emphasizing
the connection choice. In Sec.~\ref{sec:2}, we formulate the CNQ condition,
prove the no-go theorem, and study the weaker branch and its required matter
sector. The scope of the result and the perturbative calculations that remain
are discussed in Sec.~\ref{sec:limitations}. Section~\ref{sec:7} contains the
conclusions, while Appendix~\ref{app:A} supplies the connection and
minisuperspace derivations.

\section{$f(Q)$ gravity and the symmetric teleparallel connection}\label{sec:1}
In symmetric teleparallel gravity, curvature and torsion vanish identically,
and the gravitational interaction is described entirely by spacetime
non-metricity \cite{Nester:1998mp}. The fundamental geometric quantity is the
non-metricity tensor \cite{Lazkoz:2019sjl}
\begin{equation}
Q_{\lambda\mu\nu} \equiv \nabla_\lambda g_{\mu\nu},
\end{equation}
which measures the failure of the metric to remain covariantly constant.

The corresponding non-metricity scalar is constructed as \cite{BeltranJimenez:2019esp}
\begin{equation}
Q \equiv -\frac{1}{4} Q_{\lambda\mu\nu} Q^{\lambda\mu\nu}
+\frac{1}{2} Q_{\lambda\mu\nu} Q^{\nu\mu\lambda}
+\frac{1}{4} Q_\lambda Q^\lambda
-\frac{1}{2} Q_\lambda \tilde Q^\lambda,
\end{equation}
where $Q_\lambda \equiv Q_{\lambda\phantom{\mu}\mu}^{\phantom{\lambda}\mu}$ and
$\tilde Q_\lambda \equiv Q^{\mu}{}_{\lambda\mu}$.
The action of $f(Q)$ gravity is given by  \cite{BeltranJimenez:2019tme}
\begin{equation}
S =- \frac{1}{2}\int d^4x\,\sqrt{-g}\, f(Q) + S_{\mathrm{m}},
\label{fQ_action}
\end{equation}
Here $f(Q)$ is an arbitrary function of the non-metricity scalar and
$S_{\mathrm{m}}$ is the matter action. For $f(Q)=Q-2\Lambda$, the theory
reduces to STEGR and is dynamically equivalent to GR up to a boundary term.
Throughout the paper, we use units in which $8\pi G=1$.
We define the matter energy-momentum tensor by
\[
T_{\mu\nu}\equiv
-\frac{2}{\sqrt{-g}}
\frac{\delta S_{\mathrm {m}}}{\delta g^{\mu\nu}},\quad \mbox{so that} \quad  
\delta S_{\mathrm {m}}=
-\frac{1}{2}\int d^4x\,\sqrt{-g}\,
T_{\mu\nu}\,\delta g^{\mu\nu}=
\frac{1}{2}\int d^4x\,\sqrt{-g}\,
T^{\mu\nu}\,\delta g_{\mu\nu}.
\]
For the FLRW metric with lapse,
\[
ds^2=-N^2(t)\,dt^2+a^2(t)\gamma_{ij}dx^i dx^j, \quad \mbox{and a homogeneous perfect fluid} \quad 
T^\mu{}_\nu=\operatorname{diag}(-\rho,p,p,p),
\]
variation with respect to the lapse gives
\[
\frac{\delta S_{\mathrm {m}}}{\delta N}
=-a^3\rho, \quad \mbox{whereas variation with respect to the scale factor gives} \quad  
\frac{\delta S_{\mathrm {m}}}{\delta a}
=3Na^2p.
\]
A convenient background representation of the matter action is therefore
\[
S_{\mathrm {m}}^{(0)}=
-\int dt\,Na^3\rho(a), \quad \mbox{for which} \quad 
\frac{\delta S_{\mathrm {m}}^{(0)}}{\delta N}
=-a^3\rho,
\qquad
\frac{\delta S_{\mathrm {m}}^{(0)}}{\delta a}
=-Na^2\left(3\rho+a\frac{d\rho}{da}\right)
=3Na^2p,
\]
provided the usual conservation relation
\[
p=-\rho-\frac{a}{3}\frac{d\rho}{da} \quad \mbox{is used.}
\]
Thus the signs of the reduced matter terms follow directly from the covariant definition of \(T_{\mu\nu}\), rather than being imposed independently at the minisuperspace level.

A useful feature of symmetric teleparallel geometry is the existence of a
coordinate system, called the \emph{coincident gauge}, in which the affine
connection vanishes,
\begin{equation}\label{ga}
\Gamma^{\lambda}{}_{\mu\nu}=0.
\end{equation}
The coincident gauge fixes the coordinates and the connection components
simultaneously. It does not permit one to set the connection to zero while
retaining an independently prescribed coordinate form of the metric. In the
standard curved-FLRW coordinates used below, the homogeneous and isotropic
connection is nontrivial for $k\neq0$. We therefore employ an explicit
FLRW-compatible connection instead of imposing Eq.~\eqref{ga} in those
coordinates.

Varying the action (\ref{fQ_action}) with respect to the metric yields the field
equations of $f(Q)$ gravity,
\begin{equation}
\frac{2}{\sqrt{-g}} \nabla_\lambda
\left(\sqrt{-g} f_Q P^{\lambda}{}_{\mu\nu}\right)
+\frac{1}{2} g_{\mu\nu} f
+ f_Q \left(
P_{\mu\lambda\rho} Q_{\nu}{}^{\lambda\rho}
-2 Q_{\lambda\rho\mu} P^{\lambda\rho}{}_{\nu}
\right)
= -T_{\mu\nu},
\end{equation}
where $f_Q \equiv df/dQ$, $T_{\mu\nu}$ is the energy-momentum tensor, and our
superpotential convention is
\begin{equation}
P^{\lambda}{}_{\mu\nu}
=-\frac14 Q^{\lambda}{}_{\mu\nu}
+\frac12 Q_{(\mu\nu)}{}^{\lambda}
+\frac14\left(Q^{\lambda}-\widetilde Q^{\lambda}\right)g_{\mu\nu}
-\frac14\delta^{\lambda}{}_{(\mu}Q_{\nu)} .
\label{eq:superpotential}
\end{equation}
We assume minimally coupled matter with vanishing hypermomentum, so the matter
sector does not source the independent affine-connection equation.

For a homogeneous and isotropic background, $Q$ depends explicitly on the
Hubble parameter and, when spatial curvature is present, on additional
geometric structures. This feature distinguishes $f(Q)$ cosmology from its
$f(T)$ counterpart. In particular, curvature enters the $f(Q)$ background
equations through several independent terms, which is the origin of the
obstruction studied in the next section.

\section{The curvature-neutrality (CNQ) condition in $f(Q)$ gravity}\label{sec:2}

In \(f(T)\) cosmology, the flat-universe condition \(FU_T\) is implemented by
isolating the explicit spatial-curvature contribution to the modified pressure
equation and requiring its coefficient to vanish
\cite{ElHanafy:2015jbo}. This construction is possible because the curvature
dependence is carried by a single structure proportional to \(k/a^{2}\).

In \(f(Q)\) gravity, two distinct complications arise. First, spatial curvature
enters the field equations through more than one geometric structure, so it
cannot generally be eliminated by setting a single coefficient to zero.
Second, for \(k\neq0\), the coincident gauge cannot be imposed independently
of the coordinate system in which the metric is written in its standard
curved-FLRW form. We address the connection issue first because the subsequent
structural argument depends on it.

Consider the curved Friedmann-Lemaître-Robertson-Walker (FLRW) line element
\begin{equation}
ds^{2}
=
-dt^{2}
+a^{2}(t)\left[
\frac{dr^{2}}{1-kr^{2}}
+r^{2}\left(
d\theta^{2}+\sin^{2}\theta\,d\phi^{2}
\right)
\right].
\label{FLRW}
\end{equation}
Here \(a(t)\) is the scale factor, while \(k=0,+1,-1\) corresponds,
respectively, to flat, closed, and open spatial sections. In what follows, we
restrict the analysis to \(k<0\) and define
\begin{equation}
K\equiv\sqrt{-k}>0.
\label{eq:Kdef}
\end{equation}
The geometric origin of this restriction is explained below.

The coincident-gauge condition
\(\Gamma^{\lambda}{}_{\mu\nu}=0\), introduced in Eq.~\eqref{ga}, and the
coordinate representation of the metric in Eq.~\eqref{FLRW} are not
independent choices. Setting the affine connection to zero directly in the
curvature-normalised spherical coordinates of Eq.~\eqref{FLRW} does define a
flat and torsion-free connection. However, the resulting connection is not
invariant under the isometry group of the \(k\neq0\) spatial sections, and the
associated nonmetricity scalar is therefore not spatially homogeneous. A
direct calculation gives
\begin{equation}
Q_{\Gamma=0}^{(\mathrm{sph.\ FLRW})}
=
-6H^{2}
-\frac{2k}{a^{2}}
+\frac{2}{r^{2}a^{2}},
\label{eq:Qnaive}
\end{equation}
where \(H\equiv\dot a/a\). The explicit dependence on the radial coordinate
\(r\) shows that the metric-connection pair leading to
Eq.~\eqref{eq:Qnaive} does not describe a homogeneous and isotropic
background.

Consequently, for \(k\neq0\), imposing the coincident gauge in the coordinates
of Eq.~\eqref{FLRW} is not an independent coordinate choice. Once the metric
coordinates have been fixed, the condition
\(\Gamma^{\lambda}{}_{\mu\nu}=0\) selects a particular flat connection, and
that connection is not compatible with the spatial symmetries of curved FLRW
spacetime. Homogeneous and isotropic flat, torsion-free connections have been
classified in
Refs.~\cite{BeltranJimenez:2019tme,Hohmann:2021ast,DAmbrosio:2021zpm}, and we
employ the corresponding classification here.

For \(k<0\), an FLRW-compatible flat and torsion-free connection can be
specified most conveniently through its affine coordinates, namely, the
coordinates \(X^{A}\) in which its components vanish. Introducing the radial
coordinate \(\chi\), the metric in Eq.~\eqref{FLRW} may be written equivalently
as
\begin{equation}
ds^{2}
=
-dt^{2}
+a^{2}(t)\left[
d\chi^{2}
+\frac{\sinh^{2}(K\chi)}{K^{2}}\,d\Omega^{2}
\right].
\label{eq:FLRWchi}
\end{equation}
The corresponding affine coordinates may be chosen as
\begin{equation}
X^{0}
=
\tau(t)\cosh(K\chi),
\qquad
X^{i}
=
\tau(t)\sinh(K\chi)\,n^{i}(\theta,\phi),
\label{eq:affinecoords}
\end{equation}
where \(n^{i}\) is the radial unit vector and \(\tau(t)\) is a free function
characterising the connection. The connection constructed from
Eq.~\eqref{eq:affinecoords} is flat and torsion-free by construction and is
invariant under the \(SO(3,1)\) isometry group of the hyperbolic spatial
slices. The detailed derivation, including the explicit verification that
\begin{equation}
R^{\rho}{}_{\sigma\mu\nu}=0,
\qquad
T^{\rho}{}_{\mu\nu}=0,
\label{eq:flat-torsionless}
\end{equation}
is presented in Appendix~\ref{app:A}.

Variation of the action with respect to \(\tau\) yields the connection
equation of motion. As shown in Appendix~\ref{app:A}, this equation admits,
for sufficiently differentiable \(f\), the solution branch
\begin{equation}
\frac{\dot\tau}{\tau}
=
-\delta\,\frac{K}{a},
\qquad
\delta=\pm1.
\label{eq:branch}
\end{equation}
On the branch defined by Eq.~\eqref{eq:branch}, the nonmetricity scalar reduces
to the homogeneous expression
\begin{equation}
Q_{\mathrm{branch}}
=
6\left(
H+\delta\,\frac{\sqrt{-k}}{a}
\right)^{2}.
\label{QcoincidentFLU}
\end{equation}

The derivations of Eqs.~\eqref{eq:Qnaive} and
\eqref{QcoincidentFLU} both involve flat and torsion-free connections, each of
which can be made to vanish in an appropriate coordinate system. However, the
coordinate system in which the connection vanishes is different in the two
derivations.
Equation~\eqref{eq:Qnaive} is obtained by imposing
\begin{equation}
\Gamma^{\lambda}{}_{\mu\nu}(t,r,\theta,\phi)=0
\label{eq:naive-coincident-condition}
\end{equation}
directly in the spherical cosmological coordinates in which the metric takes
the standard FLRW form~\eqref{FLRW}. This choice retains the metric
components of Eq.~\eqref{FLRW} while declaring the connection components to
vanish in those same coordinates. The resulting metric-connection pair yields
Eq.~\eqref{eq:Qnaive}, whose explicit \(r\)-dependence shows that this
connection is not compatible with the homogeneity and isotropy of the
\(k\neq0\) spatial sections.

By contrast, the FLRW-compatible connection is defined by requiring its
components to vanish in the affine coordinates \(X^{A}\) introduced in
Eq.~\eqref{eq:affinecoords},
\begin{equation}
\Gamma^{A}{}_{BC}(X)=0.
\label{eq:affine-coincident-condition}
\end{equation}
When this connection is expressed in the cosmological coordinates
\(x^{\mu}=(t,\chi,\theta,\phi)\), in which the metric takes the FLRW form
\eqref{eq:FLRWchi}, its components are
\begin{equation}
\Gamma^{\lambda}{}_{\mu\nu}(x)
=
\frac{\partial x^{\lambda}}{\partial X^{A}}
\frac{\partial^{2}X^{A}}
     {\partial x^{\mu}\partial x^{\nu}},
\label{eq:transformed-connection}
\end{equation}
and are generally nonzero. Thus, in the cosmological coordinates,
\begin{equation}
\Gamma^{\lambda}{}_{\mu\nu}(t,\chi,\theta,\phi)\neq0,
\end{equation}
even though the same connection vanishes in the affine coordinates \(X^{A}\).
The metric in the affine coordinates is correspondingly the coordinate
transform of Eq.~\eqref{eq:FLRWchi}; it does not retain the standard
spherical FLRW component form.

After the connection defined by Eqs.~\eqref{eq:affinecoords} and
\eqref{eq:transformed-connection} is combined with the FLRW metric
\eqref{eq:FLRWchi}, and the solution branch~\eqref{eq:branch} of the
connection field equation is imposed, the nonmetricity scalar becomes
\begin{equation}
Q_{\mathrm{branch}}
=
6\left(
H+\delta\,\frac{\sqrt{-k}}{a}
\right)^{2}
=
6H^{2}
+12\delta H\frac{\sqrt{-k}}{a}
-6\frac{k}{a^{2}}.
\label{eq:Qbranch-expanded}
\end{equation}
Equations~\eqref{eq:Qnaive} and \eqref{eq:Qbranch-expanded} therefore do not
represent two evaluations of \(Q\) for the same metric components and the same
connection components. Equation~\eqref{eq:Qnaive} uses a connection that is
artificially set to zero in the spherical FLRW coordinates, whereas
Eq.~\eqref{eq:Qbranch-expanded} uses a connection that vanishes in the affine
coordinates \(X^{A}\) and becomes nonzero when transformed to the cosmological
coordinates. The difference between the two results is therefore a difference
between the underlying metric-connection pairs, not an algebraic
inconsistency.

The branch in Eq.~\eqref{eq:branch}, and hence the explicitly real
parametrisation in Eq.~\eqref{QcoincidentFLU}, is defined here for \(k<0\).
This is the origin of the restriction stated above. The closed case requires
a different branch of the homogeneous and isotropic connection classification
and is not considered in the present analysis. Along the branch
\eqref{eq:branch}, the limit \(k\to0\) gives
\begin{equation}
\dot\tau\longrightarrow0,
\qquad
Q_{\mathrm{branch}}\longrightarrow6H^{2}.
\label{eq:flat-limit}
\end{equation}
After the constant affine rescaling required to obtain a nonsingular
\(K\to0\) limit of the spatial affine coordinates, Eq.~\eqref{eq:flat-limit}
reproduces the usual spatially flat result. Importantly, this limiting
procedure does not amount to imposing a vanishing connection in spherical
FLRW coordinates.

It is convenient to introduce
\begin{equation}
x
\equiv
H+\delta\,\frac{\sqrt{-k}}{a},
\qquad
Q_{\mathrm{branch}}=6x^{2}.
\label{eq:xdef}
\end{equation}
The expansion in Eq.~\eqref{eq:Qbranch-expanded} shows that spatial curvature
enters through the two structures
\begin{equation}
\frac{k}{a^{2}},
\qquad
\frac{H\sqrt{-k}}{a}.
\label{eq:curvature-structures}
\end{equation}
These structures occur in the fixed combination \(x\), but generally enter
the energy and pressure equations with different weights. Additional
curvature dependence also arises through \(\dot Q\). Consequently, the
curvature dependence cannot be factorised into a single coefficient
multiplying \(k/a^{2}\), as required by the corresponding \(f(T)\)
construction. This constitutes the structural obstruction considered below.

We define the strong CNQ condition by requiring the homogeneous background
dynamics to be independent of spatial curvature for an arbitrary scale factor
\(a(t)\).\footnote{Here, ``background'' denotes the homogeneous and isotropic
FLRW solution before perturbations are introduced.} On the connection branch
\eqref{eq:branch}, this condition requires every explicitly
curvature-dependent contribution to the background field equations to vanish.

The metric field equations of \(f(Q)\) gravity may be written in a
Friedmann-like form by defining an effective energy-momentum tensor,
\begin{equation}
T^{\mu}{}_{\nu}
=
\operatorname{diag}\left(
-\rho_{\mathrm{eff}},
p_{\mathrm{eff}},
p_{\mathrm{eff}},
p_{\mathrm{eff}}
\right).
\label{eq:effective-EMT}
\end{equation}
Here \(\rho_{\mathrm{eff}}\) and \(p_{\mathrm{eff}}\) denote the source
quantities required by the gravitational field equations; they do not
represent an additional fluid introduced alongside the matter sector. When an
explicit matter interpretation is adopted in
Sec.~\ref{subsec:matter}, we identify
\begin{equation}
\rho_{\mathrm m}
=
\rho_{\mathrm{eff}},
\qquad
p_{\mathrm m}
=
p_{\mathrm{eff}}.
\label{eq:source_identification}
\end{equation}

For convenience, we multiply the complete reduced action implied by
Eq.~\eqref{fQ_action} and $S_{\mathrm m}^{(0)}$ by the irrelevant overall
factor $-1$. We therefore vary the equivalent minisuperspace action
\begin{equation}
S=\int dt\left[\tfrac{1}{2}\,N a^{3} f\big(Q(N,a,\dot a,\tau,\dot\tau,\ddot\tau)\big)
+N a^{3}\rho(a)\right]
\label{eq:minisuper}
\end{equation}
with respect to $N$, $a$ and $\tau$, setting $N=1$ after the variation and
evaluating on the branch \eqref{eq:branch}. The derivation is given in
Appendix~\ref{app:A}. One finds the effective energy density
\begin{equation}
\rho_{\mathrm{eff}}
= -\frac{1}{2} f
+ 6 H x\, f_Q ,
\label{rho_eff}
\end{equation}
and the effective pressure
\begin{equation}
p_{\mathrm{eff}}
= \frac{1}{2} f
- 2\left(\dot x + x^{2}+2Hx\right) f_Q
- 2 x\, \dot Q\, f_{QQ} ,
\label{p_eff}
\end{equation}
with $x$ defined in Eq.~\eqref{eq:xdef} and
$\dot x=\dot H-\delta\sqrt{-k}\,H/a$.

Two properties of Eqs.~\eqref{rho_eff} and \eqref{p_eff} are essential for the
subsequent argument.

First, the energy equation contains no $f_{QQ}$ term, in agreement with the
algebraic dependence of the first $f(Q)$ Friedmann equation on $f$ and $f_Q$.
Second, the pressure equation is first order in $\dot H$ and contains neither
$f_{QQQ}$ nor $\ddot Q$. The background system is therefore second order in
$a(t)$.

Both equations reproduce the required limits, which we record explicitly. For
$k\to0$ one has $x\to H$, $Q\to6H^{2}$, and
\begin{equation}
\rho_{\mathrm{eff}}\to 6H^{2}f_Q-\tfrac12 f,
\qquad
p_{\mathrm{eff}}\to \tfrac12 f-2\big(\dot H+3H^{2}\big)f_Q-24H^{2}\dot H f_{QQ},
\end{equation}
which are the standard spatially flat $f(Q)$ equations; eliminating $f$ between
them gives $\big(12H^{2}f_{QQ}+f_Q\big)\dot H=-\tfrac12(\rho+p)$. For
$f(Q)=Q$, so that $f_Q=1$ and $f_{QQ}=0$, Eqs.~\eqref{rho_eff} and
\eqref{p_eff} give
\begin{equation}
\rho_{\mathrm{eff}}=3H^{2}+\frac{3k}{a^{2}},
\qquad
p_{\mathrm{eff}}=-\left(3H^{2}+2\dot H\right)-\frac{k}{a^{2}},
\end{equation}
which are the Friedmann equations of general relativity with spatial curvature,
as they must be, STEGR being dynamically equivalent to general relativity. We
have verified both limits by direct substitution.

Equations \eqref{rho_eff} and \eqref{p_eff} show that spatial curvature
contributes through the combination $x$ in the energy equation and through $x$,
$\dot x$ and $\dot Q$ in the pressure equation, with different weights.

\subsection{The strong CNQ condition and the no-go theorem}
\label{subsec:nogo}

We now state the central result. Let $K=\sqrt{-k}$ and regard
$\rho_{\mathrm{eff}}$ and $p_{\mathrm{eff}}$ of Eqs.~\eqref{rho_eff} and
\eqref{p_eff} as functions of $K$ at fixed $a(t)$. Although FLRW curvature is
often normalised to $k=-1,0,+1$, keeping $K$ continuous is equivalent to fixing
the comoving radial coordinate convention while varying the physical curvature
scale $K/a$. Thus the derivative below compares geometries at fixed scale
factor and its complete time history (and hence $H$, $\dot H$, and higher time
derivatives) fixed, rather than differentiating a coordinate label.
Equivalently, after fixing the comoving radial-coordinate convention, the
derivative varies the physical curvature scale $K/a$. This is a deliberately
strong background-equation definition; it is not a claim of observational
degeneracy, equality of solution spaces, or invariance under a simultaneous
rescaling of $k$ and $a$.

\medskip
\noindent\textbf{Definition (strong CNQ).}
A three-times continuously differentiable function
$f\in C^3((0,\infty))$ satisfies the strong CNQ condition on the
branch \eqref{eq:branch} if
\begin{equation}
\frac{\partial \rho_{\mathrm{eff}}}{\partial K}=0
\qquad\text{and}\qquad
\frac{\partial p_{\mathrm{eff}}}{\partial K}=0
\label{eq:strongCNQ}
\end{equation}
identically, that is, for every scale factor $a(t)$ and every $K>0$.

\medskip
\noindent\textbf{Theorem.}
\emph{Within the open-FLRW branch \eqref{eq:branch}, the only
$f\in C^3((0,\infty))$ satisfying the strong CNQ condition is
$f=\mathrm{const}$.}

\medskip
\noindent\textbf{Proof.}
Since $f$, $f_Q$ and $f_{QQ}$ depend on $K$ only through $Q=6x^{2}$, and
$\partial x/\partial K=\delta/a$, so that $\partial Q/\partial K=12\delta x/a$,
differentiation of Eq.~\eqref{rho_eff} gives
\begin{equation}
\frac{\partial \rho_{\mathrm{eff}}}{\partial K}
=\frac{6\delta}{a}\left[12\,H x^{2} f_{QQ}-\frac{\delta K}{a}\,f_Q\right].
\label{eq:drhodK}
\end{equation}
Note that the $f$ and $f_Q$ variations combine so that no $f_{QQQ}$ survives.
Evaluate Eq.~\eqref{eq:drhodK} in the limit $K\to0$ at fixed $H\neq0$, where
$x\to H$:
\begin{equation}
\frac{\partial \rho_{\mathrm{eff}}}{\partial K}\bigg|_{K\to0}
=\frac{72\,\delta}{a}\,H^{3} f_{QQ}\big(6H^{2}\big).
\end{equation}
Requiring this to vanish for arbitrary $H$ forces $f_{QQ}(Q)=0$ for all $Q>0$.
Inserting $f_{QQ}=0$ back into Eq.~\eqref{eq:drhodK} leaves
\begin{equation}
\frac{\partial \rho_{\mathrm{eff}}}{\partial K}
=-\frac{6K}{a^{2}}\,f_Q ,
\end{equation}
whose vanishing for $K>0$ forces $f_Q=0$. Hence $f=\mathrm{const}$. Such an
action contains no metric kinetic term (its constant term is
cosmological-constant-like), and therefore no nondegenerate gravitational
dynamics. The pressure condition is then satisfied
trivially. %$\blacksquare$

\medskip
The result is stronger than a reduction to STEGR. To see the distinction,
substitute the linear
Lagrangian $f=\alpha Q+\beta$ into Eq.~\eqref{rho_eff} gives
\begin{equation}
\rho_{\mathrm{eff}}
=-3\alpha x^{2}+6\alpha H x-\frac{\beta}{2}
=-\frac{\beta}{2}+3\alpha H^{2}+\frac{3\alpha k}{a^{2}},
\label{eq:linearlagr}
\end{equation}
where the terms linear in $\sqrt{-k}$ have cancelled between $-f/2$ and
$6Hxf_Q$, and we have used $(\delta\sqrt{-k}/a)^{2}=-k/a^{2}$. What survives is
exactly the general-relativistic curvature term, in agreement with the STEGR
limit recorded above. Equation \eqref{eq:linearlagr} is
manifestly $k$ dependent. STEGR is therefore not curvature-neutral, as the GR
Friedmann equation itself contains $3k/a^{2}$. Thus, strong curvature
neutrality does not select STEGR. No $f(Q)$ theory in this class, including
GR, can hide spatial curvature at the background level. In the sense of
Eq.~\eqref{eq:strongCNQ}, the condition is empty.

\subsection{Weak, background-level curvature cancellation}
\label{subsec:weak}

A weaker possibility is to require cancellation only on a specified
background. We therefore assume that the curvature and expansion scales
remain in a fixed ratio,
\begin{equation}
\frac{\delta\sqrt{-k}}{a}=c\,H,
\qquad c=\mathrm{const},
\label{eq:cdef}
\end{equation}
so that $x=(1+c)H$ and $Q=6(1+c)^{2}H^{2}$. Equation \eqref{eq:cdef} integrates
immediately to $\delta\sqrt{-k}\,a^{-1}=c\,\dot a/a$, that is $\dot a=\delta\sqrt{-k}/c$,
so \emph{any} such background is coasting,
\begin{equation}
a(t)=a_{0}+\frac{\delta\sqrt{-k}}{c}\,t,
\qquad \ddot a =0,
\qquad \dot H=-H^{2}.
\label{eq:coasting}
\end{equation}
Hence, linear expansion is not independently predicted by the construction;
it follows directly from the assumed constant ratio of curvature and Hubble
scales.

In this subsection we take $f\in C^3((0,\infty))$, since differentiating the
pressure equation with respect to $K$ introduces $f_{QQQ}$.

Imposing $\partial\rho_{\mathrm{eff}}/\partial K=0$ on this background,
Eq.~\eqref{eq:drhodK} with $12H x^{2}=2QH$ and $\delta K/a=cH$ gives the single
functional condition
\begin{equation}
2Q\,f_{QQ}-c\,f_Q = 0 .
\label{eq:weakcond}
\end{equation}
The corresponding condition from the pressure equation reads
$4Q^{2}f_{QQQ}+2Q(1-c)f_{QQ}+c\,f_Q=0$, and is \emph{not} an independent
requirement: differentiating Eq.~\eqref{eq:weakcond} gives
$2Qf_{QQQ}=(c-2)f_{QQ}$, and substituting this reduces the pressure condition to
$-\big(2Qf_{QQ}-c f_Q\big)=0$. The two field equations therefore impose one and
the same constraint, which is a non-trivial consistency check on
Eqs.~\eqref{rho_eff} and \eqref{p_eff}.

Equation \eqref{eq:weakcond} integrates to $f_Q\propto Q^{c/2}$, and hence to
\begin{equation}
f(Q)= A\,Q^{(c+2)/2}+B \quad (c\neq-2),
\qquad
f(Q)= A\ln Q+B \quad (c=-2),
\label{eq:fgeneral}
\end{equation}
the logarithmic case arising because $c=-2$ is precisely the value for which the
primitive of $Q^{c/2}$ degenerates. Note that $c=-2$ is also the value for which
$x=(1+c)H=-H$ and $Q=6H^{2}$, so the non-metricity scalar on that background
coincides numerically with its spatially flat value; we record the branch for
completeness and do not pursue it further.

The weak condition therefore selects no unique Lagrangian. It defines a
one-parameter family labelled by the ratio $c$, which remains undetermined.
Cancelling one combination of curvature terms on one background is much
weaker than genuine curvature neutrality, and the resulting solution is not
curvature-neutral in an invariant sense.

For definiteness we follow the choice $c=-3$ made in the $f(T)$ literature,
\begin{equation}\label{HFLU}
H = -\frac{\sqrt{-k}}{3\delta\,a},
\qquad
a(t) = a_0 - \frac{\sqrt{-k}}{3\delta}\, t ,
\end{equation}
for which $x=-2H$ and Eq.~\eqref{eq:weakcond} becomes
\begin{equation}
2Q f_{QQ} + 3 f_Q = 0 ,
\label{eq:weak3}
\end{equation}
with general solution
\begin{equation}
f(Q) = \frac{\alpha_1}{\sqrt{Q}} + \beta_1 .
\label{eq:fsol}
\end{equation}
Using Eq.~\eqref{QcoincidentFLU} together with \eqref{HFLU}, the non-metricity
scalar reduces to
\begin{equation}
Q=\frac{8}{3}\frac{-k}{a^2}\;\propto\;a^{-2},
\label{eq:Qback}
\end{equation}
so that
\begin{equation}
{
f(t)=A\,a(t)+\beta_1,
\qquad
f_Q \;\propto\; Q^{-3/2}\;\propto\; a^{3},
}
\label{eq:fofa}
\end{equation}
where $A$ is a constant. We impose the conventional positive-coupling sign
$f_Q>0$. Since $Q>0$ on this background, this assumption requires
$\alpha_1<0$. We stress that $f_Q>0$ is adopted here as a physical sign
condition continuously connected to the positive-coupling STEGR convention;
it is not presented as a tensor no-ghost condition on the non-flat branch,
whose quadratic action has not been derived.

\subsection{Physical interpretation}

The weak branch describes a coasting universe with $\ddot a=0$ and $q=0$.
Its effective fluid is obtained by evaluating Eqs.~\eqref{rho_eff} and
\eqref{p_eff} on \eqref{HFLU} and \eqref{eq:fsol}:
\begin{equation}
\rho_{\mathrm{eff}}(t)=\rho_0\,a(t)+\rho_{\Lambda},
\qquad
p_{\mathrm{eff}}(t)=-\frac{4}{3}\rho_0\,a(t)-\rho_{\Lambda},
\label{eq:fluid}
\end{equation}
where $\rho_{\Lambda}=-\beta_1/2$ is an effective cosmological constant. We have
checked that Eqs.~\eqref{eq:fluid} satisfy
$\dot\rho_{\mathrm{eff}}+3H(\rho_{\mathrm{eff}}+p_{\mathrm{eff}})=0$
identically, which is an independent test of Eqs.~\eqref{rho_eff} and
\eqref{p_eff}. The component growing as $a$ corresponds to
\begin{equation}
{
w=-\frac{4}{3},
}
\end{equation}
while the constant contribution satisfies $w=-1$. If
$\rho_0>0$ and $\rho_\Lambda>0$, the total equation of state evolves from
$w_{\mathrm{eff}}\to-1$ as $a\to0$ to
$w_{\mathrm{eff}}\to-4/3$ as $a\to\infty$. For other signs of
$\rho_\Lambda$, this interpolation need not hold and
$\rho_{\mathrm{eff}}$ may cross zero. For $\rho_0>0$, the growing component is
of phantom type. We regard this behavior as a defect of the construction and
do not present it as a viable dark-energy model.

The geometry of this background requires clarification. Although the solution
\eqref{eq:coasting} is coasting, the choice $c=-3$ does not describe the Milne
universe. For a general coasting FLRW metric, the Ricci scalar is
\begin{equation}
R=\frac{6\left(\dot a^{2}+k\right)}{a^{2}}
=\frac{6k}{a^{2}}\left(1-\frac{1}{c^{2}}\right),
\label{eq:Ricci}
\end{equation}
where the second equality uses $\dot a=\delta\sqrt{-k}/c$ from
Eq.~\eqref{eq:coasting}. The spacetime is therefore flat if and only if
$c^{2}=1$. For $c=-3$ one has $R=\tfrac{16}{3}\,k/a^{2}\neq0$, which for $k=-1$
and $a=t/3$ reads $R=-48/t^{2}$. The $c=-3$ background is genuinely curved and
is not locally isometric to Minkowski space.

The Milne case $c^{2}=1$ is, on the expanding branch $\delta=-1$ considered
here, the value $c=-1$, and it is degenerate. There $x=(1+c)H$ vanishes
identically, so $Q\equiv0$ on the background. This point lies outside the
domain $Q\in(0,\infty)$ assumed in deriving Eq.~\eqref{eq:fgeneral}. Formally
continuing that family gives $f\propto Q^{1/2}$ and
$f_Q\propto Q^{-1/2}$, which is singular at $Q=0$; it is therefore not a
regular member of the weak family. The vanishing of $Q$ is expected because
Milne is Minkowski spacetime in hyperbolic slicing and the branch
\eqref{eq:branch} reduces there to the flat connection. It nevertheless means
that the weak construction has no well-defined content at the one background
on which the spacetime curvature actually vanishes.

The valid interpretation is therefore weaker. The condition forces
$a\propto t$, the same expansion law produced by curvature in an empty open
universe, although the resulting spacetime is not that universe. This outcome
further shows that the weak condition does not achieve physical curvature
neutrality.

\begin{figure*}[t]
\centering
\includegraphics[width=0.98\textwidth]{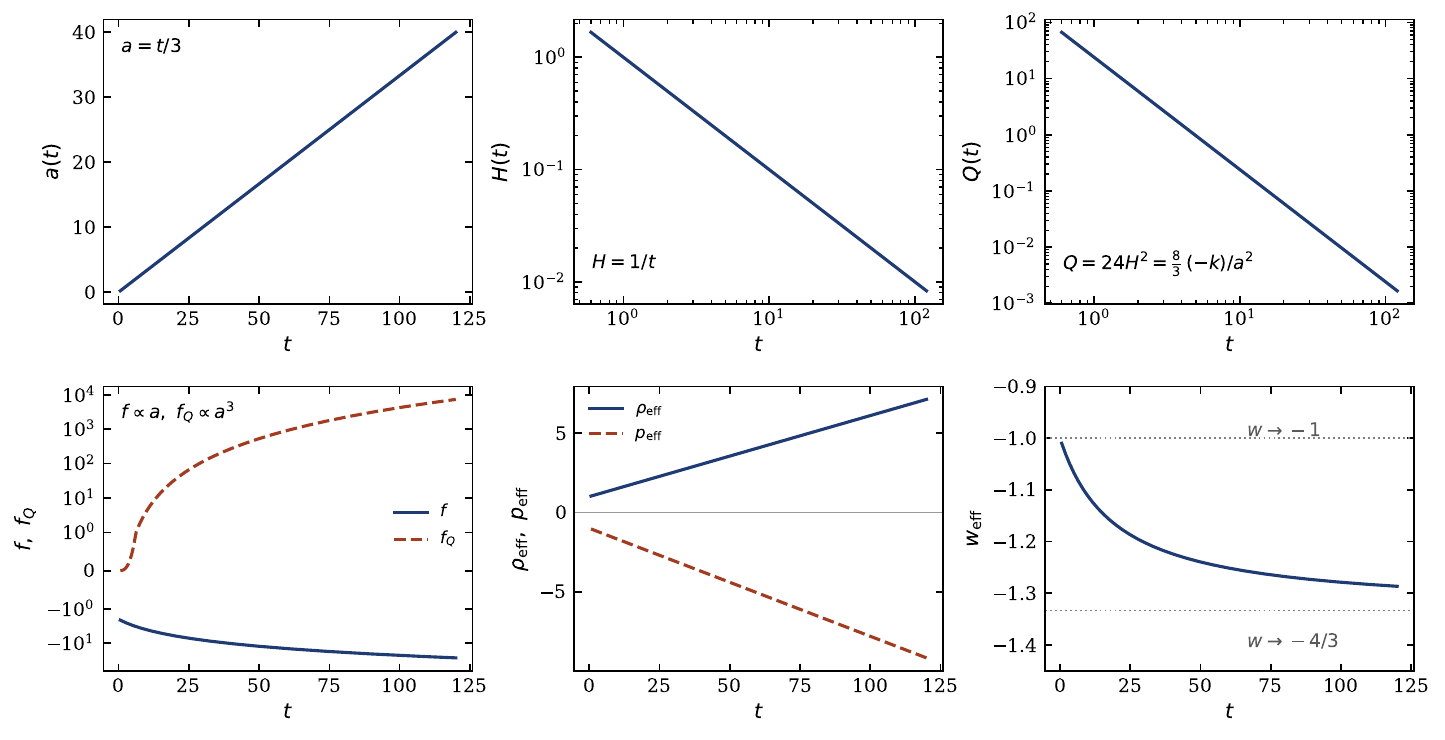}
\caption{The $c=-3$ weak-CNQ background on the expanding branch $\delta=-1$,
for $k=-1$, $a_{0}=0$, $\alpha_1=-1$ and $\beta_1=-2$ (so that
$\rho_{\Lambda}=1$). A reference time $t_\star$ is used to make all plotted
quantities dimensionless: the axes represent $t/t_\star$, $a/t_\star$,
$Ht_\star$, $Qt_\star^2$, $ft_\star^2$, and $f_Q$ in the corresponding
reference units. Panels show the scale factor $a=t/3$, the Hubble rate
$H=1/t$, the non-metricity scalar $Q=24H^{2}$, the Lagrangian
$f=\alpha_1/\sqrt{Q}+\beta_1$, the effective density
$\rho_{\mathrm{eff}}=\rho_0 a+\rho_{\Lambda}$ and the effective equation of
state $w_{\mathrm{eff}}=p_{\mathrm{eff}}/\rho_{\mathrm{eff}}$. The dashed lines
in the last panel mark the asymptotic values $w=-1$ at early times and
$w=-4/3$ at late times. The growing component is of phantom type; we report it
because it is what the equations give, not as a viable dark-energy model.}
\label{fig:background}
\end{figure*}

The deceleration parameter vanishes identically on this background, $q=0$,
consistent with $\ddot a=0$. The background quantities of this subsection are
collected in Fig.~\ref{fig:background}.

\subsection{The matter sector required by the weak condition}
\label{subsec:matter}

So far, the source required by the weak branch has been represented by one
effective fluid. We now determine whether it can be decomposed into separately
conserved radiation, pressureless matter, and a residual component.

Equation \eqref{eq:cdef} fixes $H(a)$ kinematically and independently of the
source. Once matter is introduced, the imposed expansion law and Friedmann
constraint must hold simultaneously. This does not forbid a source; instead,
it determines the source uniquely. Writing the total matter content
as $(\rho_{\mathrm m},p_{\mathrm m})$, the field equations on the background
\eqref{HFLU} with the Lagrangian \eqref{eq:fsol} read
\begin{equation}
\rho_{\mathrm m}(a)=\rho_0\,a+\rho_{\Lambda},
\qquad
p_{\mathrm m}(a)=-\frac{4}{3}\rho_0\,a-\rho_{\Lambda},
\label{eq:required}
\end{equation}
where
\begin{equation}
\rho_0=-\frac{\sqrt{6}\,\alpha_1}{16\sqrt{-k}}.
\label{eq:rho0}
\end{equation}
Thus, Eq.~\eqref{eq:required} is not an assumed equation of state but a direct
consequence of the weak CNQ condition.

Under the conventional positive-coupling assumption $f_Q>0$, which implies
$\alpha_1<0$ and hence $\rho_0>0$, the required source has two immediate
properties. First, it violates the null energy condition at all times:
\begin{equation}
\rho_{\mathrm m}+p_{\mathrm m}
=-\frac{1}{3}\rho_0\,a<0 .
\label{eq:NEC}
\end{equation}
The magnitude of this violation increases monotonically as the universe
expands. This conclusion relies on the physically selected sign $f_Q>0$ and
is therefore not a sign-independent consequence of the weak CNQ condition
alone. It is a statement about the total content and is independent of how one
chooses to split it.

\begin{figure}[t]
\centering
\includegraphics[width=0.47\textwidth]{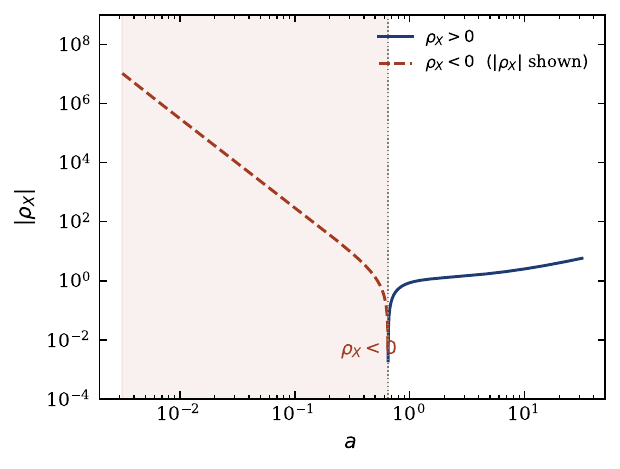}
\caption{The residual component $\rho_X$ of Eq.~\eqref{eq:rhoX} required to
support the $c=-3$ background once radiation and pressureless matter are
included, for $\rho_{\mathrm r0}=10^{-4}$, $\rho_{\mathrm d0}=0.3$,
$\rho_{\Lambda}=1$ and $\rho_0=0.15$. Solid where $\rho_X>0$ and dashed where
$\rho_X<0$; the shaded region marks $\rho_X<0$. Any non-zero
$\rho_{\mathrm r0}$ or $\rho_{\mathrm d0}$ drives $\rho_X$ negative at small
$a$. The vertical dotted line marks the zero of $\rho_X$ for these parameters.}
\label{fig:matter}
\end{figure}

Second, the split cannot be arranged so that the familiar components behave
normally. Normalising $a(t_0)=1$ and writing
$\rho_{\mathrm m}=\rho_{\mathrm r0}a^{-4}+\rho_{\mathrm d0}a^{-3}+\rho_X$ for
radiation, pressureless matter and whatever else is required, Eq.~\eqref{eq:required}
gives
\begin{equation}
\rho_X(a)=\rho_0\,a+\rho_{\Lambda}-\rho_{\mathrm d0}a^{-3}-\rho_{\mathrm r0}a^{-4},
\qquad
p_X(a)=-\frac{4}{3}\rho_0\,a-\rho_{\Lambda}-\frac{1}{3}\rho_{\mathrm r0}a^{-4}.
\label{eq:rhoX}
\end{equation}
Since radiation and dust are assumed to be separately conserved, and the
total source \eqref{eq:required} satisfies
\[
\dot{\rho}_{\mathrm m}
+3H\left(\rho_{\mathrm m}+p_{\mathrm m}\right)=0
\]
identically, the residual component $X$ is separately conserved as well.
Equation~\eqref{eq:rhoX} is therefore dynamically consistent. However, for
$\rho_{\mathrm r0}>0$,
\[
\rho_X(a)\sim-\rho_{\mathrm r0}a^{-4}
\qquad (a\to0),
\]
whereas, if radiation is absent but $\rho_{\mathrm d0}>0$,
\[
\rho_X(a)\sim-\rho_{\mathrm d0}a^{-3}
\qquad (a\to0).
\]
Thus any nonzero positive radiation or dust contribution forces
$\rho_X<0$ at sufficiently early times, with a divergently negative energy
density. Moreover, $w_X=p_X/\rho_X$ becomes singular whenever $\rho_X$
crosses zero, although the physical variables $\rho_X$ and $p_X$ themselves
remain finite at the crossing. This behaviour is illustrated in
Fig.~\ref{fig:matter}.

The conclusion is again negative. For $c=-3$, no decomposition containing
positive radiation and pressureless matter can keep $X$ nonnegative throughout
the evolution. Radiation, dust, and the residual component therefore cannot
all possess nonnegative energy density for every $a>0$.

There is also an independent phenomenological warning. Eternally coasting
$R_h=ct$ cosmologies have been proposed in Ref.~\cite{Melia:2011fj} and found
to be strongly disfavoured by supernova, cosmic-chronometer, and radial-BAO
expansion data, particularly at low redshift, in
Ref.~\cite{Bilicki:2012ub}. That comparison is not a direct likelihood analysis
of the present curved branch, whose source sector differs, but it reinforces
the conclusion that the weak solution should be treated as a structural
diagnostic rather than a candidate cosmology.

The principal differences between the curvature-neutral constructions in
$f(T)$ and $f(Q)$ gravity are summarised in Table~\ref{tab:comparison}.

\begin{table}[t]
\centering
\caption{Comparison between curvature-neutral constructions in $f(T)$ and
$f(Q)$ gravity.}
\label{tab:comparison}
\begin{tabular}{l@{\hspace{1.2em}}c@{\hspace{1.2em}}c}
\hline\hline
 & $f(T)$ gravity & $f(Q)$ gravity ($k<0$ branch) \\
\hline
Geometric origin & Torsion & Non-metricity \\
Curvature structure & Single $k/a^2$ term & Multiple structures \\
CN under cited prescription & Refs.~\cite{ElHanafy:2014tzj,ElHanafy:2015jbo} & No under Eq.~\eqref{eq:strongCNQ} \\
Background-constrained CN & Not required & Required \\
Scale factor & Linear & Linear \\
Effective behavior & Coasting & Coasting (not Milne) \\
Required source & --- & NEC violation for $f_Q>0$ \\
Ordinary-matter split & --- & $\rho_X<0$ at small $a$ \\
\hline\hline
\end{tabular}
\end{table}

\section{Scope and perturbative limitations}\label{sec:limitations}

The theorem established above concerns only the background dynamics of the
open-FLRW connection branch \eqref{eq:branch}. It does not demonstrate the
stability of the weak coasting solution. Perturbing the imposed relation
$\ddot a=0$ would merely move within the family of coasting histories and
would not linearize the independent metric and connection equations.

Perturbation formulae derived in a spatially flat, metric-only coincident gauge
also cannot be imported into the curved-FLRW coordinates used here. The
background affine connection is nonzero and must be perturbed together with
the scalar, vector, and tensor metric sectors. A valid stability analysis must
derive the complete quadratic action, identify its constraints, and
diagonalize the kinetic and gradient matrices. Only then can one determine
whether this branch suffers from the ghost or strong-coupling problems found
on other cosmological branches of $f(Q)$ gravity
\cite{Gomes:2023tur}.

For the same reason, we make no claim about the tensor propagation speed, the
tensor no-ghost condition, or the gravitational-wave luminosity distance. In
particular, the scaling
$d_L^{\mathrm{GW}}/d_L^{\mathrm{EM}}=(1+z)^{3/2}$ that follows from inserting
$f_Q\propto a^3$ into a commonly used spatially flat tensor action is not an
established prediction on the present non-flat connection branch. Deriving the
covariant quadratic action with connection perturbations included is left for
future work.

\section{Conclusions}\label{sec:7}

In this work, we have investigated the curvature-neutral scenario in $f(Q)$
gravity on an explicitly specified open-FLRW connection branch. The analysis
was motivated by earlier curvature-neutral constructions in teleparallel
$f(T)$ gravity and by the continuing development of symmetric teleparallel
theories as alternatives to GR \cite{ElHanafy:2015jbo}.

We have shown that CNQ differs fundamentally from its $f(T)$ counterpart. On
the selected branch, curvature enters through explicit $k/a^2$ terms, mixed
$H\sqrt{-k}/a$ contributions, and implicitly through $Q$ and its time
derivative. These structures carry different weights in the background
equations and cannot be factorized into one curvature coefficient. The
$f(T)$ prescription therefore has no direct extension to $f(Q)$ gravity.

Imposing complete curvature independence for an arbitrary scale factor, we
proved that the only admissible Lagrangian is $f=\mathrm{const}$ and therefore
contains no gravitational dynamics. STEGR also fails because its Friedmann
equation retains the GR term $3k/a^{2}$. The no-go theorem is consequently
stronger than a reduction to STEGR: strict curvature neutrality selects no
dynamical theory in this class.

We then introduced a weaker notion of CNQ, in which the cancellation is imposed
on a fixed background rather than identically. Requiring the curvature and Hubble
scales to remain in a constant ratio $c$ makes the expansion coasting
automatically, and the cancellation condition reduces to the single functional
constraint $2Qf_{QQ}=c\,f_Q$, with solution $f(Q)=AQ^{(c+2)/2}+B$ for $c\neq-2$
and $f(Q)=A\ln Q+B$ at the degenerate value $c=-2$. The same
constraint follows from the energy and the pressure equations independently,
which is a useful consistency check, but $c$ itself is not fixed by the
construction: the weak condition selects a one-parameter family rather than a
model. For the value $c=-3$ inherited from the $f(T)$ literature one obtains
$f(Q)=\alpha_1/\sqrt{Q}+\beta_1$.

The resulting cosmological evolution can be represented by an effective fluid
whose energy density consists of a component growing as $a$ and a constant
component. The growing component has $w=-4/3$ and is of phantom type for the
positive-coupling sign. When both component densities are positive, the total
equation of state evolves from $w=-1$ to $w=-4/3$. We regard the phantom
component as a defect of the construction rather than a feature.
The background is coasting, and coasting expansion is what spatial curvature
alone produces in an empty open universe, so the construction does return the
kinematics it set out to remove. It does not, however, return the Milne
universe: the Ricci scalar of the $c$-family is
$R=6k\big(1-c^{-2}\big)/a^{2}$, which vanishes only for $c^{2}=1$; on the
expanding branch $\delta=-1$ treated here that is $c=-1$, where $x$ and hence
$Q$ vanish identically and the construction degenerates. For $c=-3$ one has
$R=16k/(3a^{2})\neq0$.

We also showed that the weak condition determines the matter sector rather than
leaving it free. The required source is unique, violates the null energy
condition at all times, and cannot be split so that radiation and pressureless
matter carry positive energy density at early epochs: any non-zero amount of
either forces a compensating component with divergently negative density as
$a\to0$. Independently, a strictly coasting expansion is incompatible with
nucleosynthesis and the CMB acoustic scale
\cite{Melia:2011fj,Bilicki:2012ub}. The weak branch is therefore not a candidate
cosmology, and we do not offer it as one.

%As detailed in Sec.~\ref{sec:limitations}, no perturbative stability or
%gravitational-wave result follows from the present background calculation.
%Those questions require the complete metric-connection quadratic action on
%the non-flat branch and are left open.

The negative result is the substantive conclusion of the study. Under the
strong definition \eqref{eq:strongCNQ}, spatial curvature cannot be hidden at
the background level on the open-FLRW branch considered here. The closest
construction is a coasting solution with
$f\propto Q^{-1/2}$, but it is not curvature-neutral in an invariant sense,
does not describe the Milne universe, and, for $f_Q>0$, requires a
NEC-violating source. Adding positive radiation or dust then drives the
residual component negative at early times. Deriving the quadratic action on
non-flat spatial sections and testing the perturbative pathologies identified
in Ref.~\cite{Gomes:2023tur}, with connection perturbations included, remain
important problems for future work.

\section*{Data Availability Statement}

No external datasets were used. All numerical values underlying the figures
follow directly from the equations and parameter choices stated in the text.

\appendix

\section{The symmetric teleparallel connection for open FLRW, and the
cosmological field equations}
\label{app:A}

This appendix supplies the derivation of Eq.~\eqref{QcoincidentFLU} and of the
field equations \eqref{rho_eff} and \eqref{p_eff}. All algebra reported here has
been carried out symbolically and cross-checked numerically.

\subsection{Why $\Gamma^{\lambda}{}_{\mu\nu}=0$ cannot be used with Eq.~\eqref{FLRW}}

Symmetric teleparallel gravity requires a connection that is flat,
$R^{\rho}{}_{\sigma\mu\nu}=0$, and torsion-free,
$T^{\rho}{}_{\mu\nu}=0$, but not metric compatible. Any such connection admits
coordinates in which its components vanish. This is the coincident gauge. The
freedom is a coordinate freedom, and it is therefore not independent of the
freedom used to write the metric in a particular form.

If one nevertheless sets $\Gamma^{\lambda}{}_{\mu\nu}=0$ in the curvature-normalised
coordinates $(t,r,\theta,\phi)$ of Eq.~\eqref{FLRW}, then
$Q_{\lambda\mu\nu}=\partial_\lambda g_{\mu\nu}$ and a direct evaluation of the
definition of $Q$ gives Eq.~\eqref{eq:Qnaive},
$Q=-6H^{2}-2k/a^{2}+2/(r^{2}a^{2})$. The $r$-dependence shows that the
connection so defined is not invariant under the isometries of the $k\neq0$
spatial sections, so this choice does not describe a homogeneous and isotropic
geometry. For $k=0$ in Cartesian coordinates the same computation gives
$Q=6H^{2}$, the familiar result, which is why the issue does not arise in the
spatially flat case. The general classification of homogeneous and isotropic
symmetric teleparallel geometries, and the observation that the cosmological and
coincident coordinate systems do not in general agree, are due to
Refs.~\cite{Hohmann:2021ast,DAmbrosio:2021zpm}.

\subsection{The $k<0$ branch}

Write the open FLRW metric as
\begin{equation}
ds^{2}=-N^{2}(t)\,dt^{2}
+a^{2}(t)\left[d\chi^{2}+\frac{\sinh^{2}(K\chi)}{K^{2}}\,d\Omega^{2}\right],
\qquad K\equiv\sqrt{-k},
\label{eq:A_metric}
\end{equation}
which is Eq.~\eqref{FLRW} after $r=\sinh(K\chi)/K$. We specify the connection by
its affine coordinates, Eq.~\eqref{eq:affinecoords},
\begin{equation}
X^{0}=\tau(t)\cosh(K\chi),
\qquad
X^{i}=\tau(t)\sinh(K\chi)\,n^{i}(\theta,\phi),
\end{equation}
so that
\begin{equation}
\Gamma^{\lambda}{}_{\mu\nu}
=\frac{\partial x^{\lambda}}{\partial X^{A}}\,
\frac{\partial^{2}X^{A}}{\partial x^{\mu}\partial x^{\nu}} .
\label{eq:A_gamma}
\end{equation}
A connection of the form \eqref{eq:A_gamma} is flat and torsion-free
identically, and we have verified both explicitly. The map is the $k<0$
counterpart of the Cartesian coincident gauge: it is the Milne foliation, whose
isometry group $SO(3,1)$ is exactly the isometry group of the hyperbolic slices,
so the connection is compatible with homogeneity and isotropy. A constant
rescaling of the $X^{i}$ leaves \eqref{eq:A_gamma} unchanged, so no factor of
$K$ need be carried in the spatial affine coordinates. The single free function
$\tau(t)$ is the additional connection degree of freedom identified in
Ref.~\cite{Hohmann:2021ast}.

Computing $Q_{\lambda\mu\nu}=\nabla_\lambda g_{\mu\nu}$ with the metric
\eqref{eq:A_metric} and the connection \eqref{eq:A_gamma}, and contracting, one
obtains a non-metricity scalar that is a function of $t$ alone. Writing
$\lambda\equiv\ln\tau$,
\begin{align}
Q =\;& \frac{6K^{2}}{a^{2}}
+\frac{3K^{2}}{a^{2}}\left[\frac{\ddot\lambda}{\dot\lambda^{2}}
-\frac{\dot a}{a\dot\lambda}-\frac{\dot N}{N\dot\lambda}\right]
\nonumber\\
&-\frac{3\ddot\lambda}{N^{2}}
-\frac{9\dot a\dot\lambda}{N^{2}a}
+\frac{6\dot a^{2}}{N^{2}a^{2}}
+\frac{3\dot N\dot\lambda}{N^{3}} .
\label{eq:A_Qgen}
\end{align}
Because Eq.~\eqref{eq:A_Qgen} contains ratios of $K$ and
$\dot\lambda$, an unconstrained simultaneous limit $K\to0$,
$\dot\lambda\to0$ is path dependent. The spatially flat result is recovered
along the solution branch below: substituting
$\dot\lambda=-\delta NK/a$ first and then taking $K\to0$ gives
$Q\to6\dot a^{2}/(N^{2}a^{2})=6H^{2}$.

\subsection{The connection equation and the branch}

Varying the minisuperspace action \eqref{eq:minisuper} with respect to $\lambda$
gives the connection equation of motion. Because $Q$ depends on $\lambda$ only
through $\dot\lambda$ and $\ddot\lambda$, this equation is a total derivative.
More explicitly, with $L_g=Na^3f(Q)/2$ it is
\begin{equation}
-\frac{d}{dt}\left(\frac{Na^3}{2}f_Q
\frac{\partial Q}{\partial\dot\lambda}\right)
+\frac{d^2}{dt^2}\left(\frac{Na^3}{2}f_Q
\frac{\partial Q}{\partial\ddot\lambda}\right)=0.
\label{eq:A_connection_EL}
\end{equation}
The derivatives entering Eq.~\eqref{eq:A_connection_EL} follow directly
from Eq.~\eqref{eq:A_Qgen}:
\begin{equation}
\frac{\partial Q}{\partial\ddot{\lambda}}
=
\frac{3K^{2}}{a^{2}\dot{\lambda}^{2}}
-\frac{3}{N^{2}},
\label{eq:A_dQddlambda}
\end{equation}
and
\begin{equation}
\frac{\partial Q}{\partial\dot{\lambda}}
=
\frac{3K^{2}}{a^{2}}
\left(
-\frac{2\ddot{\lambda}}{\dot{\lambda}^{3}}
+\frac{\dot a}{a\dot{\lambda}^{2}}
+\frac{\dot N}{N\dot{\lambda}^{2}}
\right)
-\frac{9\dot a}{N^{2}a}
+\frac{3\dot N}{N^{3}}.
\label{eq:A_dQdlambda}
\end{equation}
Equations~\eqref{eq:A_connection_EL}-\eqref{eq:A_dQdlambda} give the
connection equation entirely off shell, before any branch condition is
imposed. 
Substituting
\begin{equation}
\dot{\lambda}
=
-\delta\,\frac{NK}{a},
\qquad
\delta^{2}=1,
\label{eq:A_branch}
\end{equation}
into Eqs.~\eqref{eq:A_connection_EL}-\eqref{eq:A_dQdlambda}, while retaining
all time derivatives acting on $f_Q$, makes the connection equation vanish
identically for sufficiently differentiable $f$. Thus
Eq.~\eqref{eq:branch} is a solution of the connection field equation rather
than an additional kinematic assumption. On this branch,
Eq.~\eqref{eq:A_Qgen} reduces to
\begin{equation}
Q
=
6\left(
\frac{\dot a}{Na}
+\delta\frac{K}{a}
\right)^2
=
6\left(
H+\delta\frac{\sqrt{-k}}{a}
\right)^2,
\label{eq:A_Qbranch}
\end{equation}
which reproduces Eq.~\eqref{QcoincidentFLU}.

\subsection{Variation with respect to $N$ and $a$}

With $Q$ given by Eq.~\eqref{eq:A_Qgen}, we vary Eq.~\eqref{eq:minisuper},
\begin{equation}
\frac{\delta S}{\delta q}
=\frac{\partial L}{\partial q}
-\frac{d}{dt}\frac{\partial L}{\partial\dot q}
+\frac{d^{2}}{dt^{2}}\frac{\partial L}{\partial\ddot q},
\qquad q\in\{N,a,\lambda\},
\end{equation}
set $N=1$ after the variation, and evaluate on the branch. Defining
\begin{equation}
\rho_{\mathrm{eff}}=-\frac{1}{a^{3}}\frac{\delta S_{g}}{\delta N}\bigg|_{N=1},
\qquad
p_{\mathrm{eff}}=\frac{1}{3a^{2}}\frac{\delta S_{g}}{\delta a}\bigg|_{N=1},
\end{equation}
with the normalisation fixed by requiring $\rho_{\mathrm{eff}}=3H^{2}$ and
$p_{\mathrm{eff}}=-(3H^{2}+2\dot H)$ for $f=Q$ and $k=0$. Carrying out the
variation off shell and imposing Eq.~\eqref{eq:A_branch} only afterwards, the
lapse equation reduces to
\begin{equation}
-\frac{1}{a^{3}}\frac{\delta S_g}{\delta N}\bigg|_{N=1}
=-\frac{1}{2}f+6Hxf_Q,
\label{eq:A_lapse_result}
\end{equation}
while the scale-factor equation becomes
\begin{equation}
\frac{1}{3a^{2}}\frac{\delta S_g}{\delta a}\bigg|_{N=1}
=\frac{1}{2}f
-2\left(\dot x+x^{2}+2Hx\right)f_Q
-2x\dot Q f_{QQ}.
\label{eq:A_scale_result}  
\end{equation}
Equations~\eqref{eq:A_lapse_result} and \eqref{eq:A_scale_result} reproduce
Eqs.~\eqref{rho_eff} and \eqref{p_eff}, respectively. Three independent checks were performed:
the $k\to0$ limit reproduces the standard spatially flat $f(Q)$ system, including
$(12H^{2}f_{QQ}+f_Q)\dot H=-\tfrac12(\rho+p)$; the choice $f=Q$ reproduces the
Friedmann equations of general relativity with curvature for both $\delta=\pm1$;
and the effective fluid \eqref{eq:fluid} obtained on the weak-CNQ background
satisfies the continuity equation
$\dot\rho_{\mathrm{eff}}+3H(\rho_{\mathrm{eff}}+p_{\mathrm{eff}})=0$ identically,
providing an additional check on the algebra and on the mutual consistency of
the two background field equations.

\end{document}